\documentclass[letterpaper,twocolumn,10pt]{article}
\usepackage{usenix2019_v3}

\usepackage{tikz}
\usepackage{amsmath}

\usepackage{filecontents}

\usepackage{algorithm}
\usepackage{algpseudocode}
\usepackage{graphicx}
\usepackage{epstopdf}
\usepackage{float}
\usepackage{subfig}
\usepackage{balance}
\usepackage{comment}
\usepackage{amssymb, amsmath}
\usepackage{multirow}
\usepackage{xspace}
\usepackage{xcolor, color, soul}
\usepackage{url}
\usepackage{pifont}
\usepackage{wasysym}

\newcommand{\name}{\textit{Locate3D}\xspace}
\newcommand\todo[1]{\textcolor{black}{#1}}

\newcommand{\new}{\vskip 0.75em plus 0.15em minus .1em}

\newcommand{\squishlist}{
	\begin{list}{$\bullet$}
		{  \setlength{\leftmargin}{+0.15in}
		} }

		\newcommand{\squishend}{
	\end{list} }

\usepackage{soul,color}
\usepackage{mathtools}
\hyphenchar\font=-1
\usepackage{filecontents}

\begin{document}

\date{}

\title{\Large \bf Real-Time 3D Localization and Orientation for Large Network of Devices}
\title{\Large \bf \vspace{-0.5in} Large Network UWB Localization: Algorithms and Implementation\vspace{-0.2in}}
\title{\Large \bf \vspace{-0.5in} Fast Localization and Tracking in City-Scale UWB Networks\vspace{-0.2in}}
\author{
{\rm Nakul Garg, Irtaza Shahid, Ramanujan K Sheshadri$^{\ddagger}$, Karthikeyan Sundaresan$^{\S}$, Nirupam Roy}\\
\{nakul22,irtaza\}@umd.edu, $^\ddagger$ram.sheshadri@nokia-bell-labs.com,\\ $^{\S}$karthik@ece.gatech.edu, niruroy@umd.edu\\
University of Maryland College Park, $^{\ddagger}$Nokia Bell Labs, $^{\S}$Georgia Institute of Technology
}

\maketitle

\begin{abstract}


Localization of networked nodes is an essential problem in emerging applications, including first-responder navigation, automated manufacturing lines, vehicular and drone navigation, asset navigation and tracking, Internet of Things and 5G communication networks.
In this paper, we present \name, a novel system for peer-to-peer node localization and orientation estimation in large networks.
Unlike traditional range-only methods, \name introduces angle-of-arrival (AoA) data as an added network topology constraint.
The system solves three key challenges: it uses angles to reduce the number of measurements required by $4\times$ and jointly use range and angle data for location estimation.
We develop a spanning-tree approach for fast location updates, and to ensure the output graphs are rigid and uniquely realizable, even in occluded or weakly connected areas. \name cuts down latency by up to $75\%$ without compromising accuracy, surpassing standard range-only solutions. It has a $10.2$ meters median localization error for large-scale networks ($30,000$ nodes, $15$ anchors spread across $14km^2$) and $0.5$ meters for small-scale networks ($10$ nodes).

\end{abstract}


\section{Introduction}
\vspace{-0.1in}
A swarm of connected nodes is the underlying architecture for many emerging applications.
Miniaturized sensing modules are scattered like seeds \cite{iyer2022wind, thompson2021floating} or carried by insects \cite{iyer2020airdropping} to scale a vast region for fine grained sensor networks toward border protection \cite{bhadwal2019smart}, animal migration tracking \cite{lee2021msail}, or precision agriculture \cite{vasisht2017farmbeats}.
Flocks of drones can localize each other to fly in formation for charting inaccessible regions \cite{scott2017drone, han2015use} or for airshows \cite{droneairshow}.
A smart network of tags can enable city-scale tracking of deliveries or missing objects in real-time \cite{minoli2018ultrawideband}.
Future of the cellular  \cite{raissi2019autonomous} and vehicular networks  \cite{alam2013cooperative} can pave the path toward autonomy  \cite{wang2018networking, zheng2015reliable} and road safety \cite{wymeersch2015challenges}.
Localization of the nodes in such large networks is an essential requirement and at the same time a challenging problem when the number of nodes is large.
In this paper, we present \name, a system that is grounded on both theoretical and practical foundations, providing a reliable framework for fast peer-to-peer localization as well as orientation of nodes on a large network.
\new

\begin{figure}[t]
    \centering
    \includegraphics[width=3.2in]{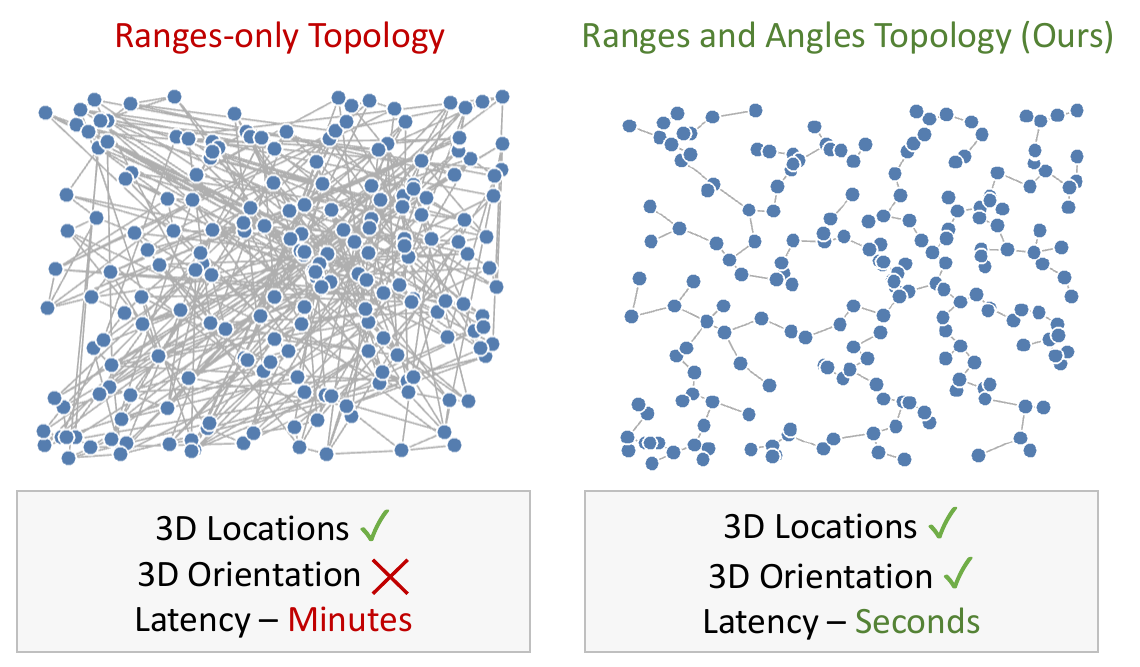}
    \caption{Adding angles to the edges between nodes introduces additional constraints to the topology, significantly decreasing the number of edges necessary to achieve a \textit{rigid} and \textit{unique} graph realization. Furthermore, this provides more information to accurately estimate the relative 3D orientations between nodes.}
    \label{fig:graphs_comparison_example}
\end{figure}

Multidimensional scaling (MDS) has been the essence of nearly all large-scale localization algorithms for wireless networks.
The primary reason MDS is adapted so widely is its ability to reconstruct the relative map of the network nodes with little infrastructural support, even in the form of the anchor nodes with known locations.
MDS-based algorithms are fundamentally centralized ranging-based systems that consider inter-node distances in the Euclidean space to optimize for sensor locations.
It, however, suited the capacity of the mobile nodes that can use the time of flight of the signal or RSSI-based model to estimate ranges without additional hardware requirements or computational complexity.
However, recent mobile nodes have evolved not only to accurately sense the ranges but are also equipped with multi-dimensional antenna arrays to estimate reliable angles of the nodes.
For instance, around 20 years after being released for commercial applications in 2002 \cite{firacons}, off-the-shelf UWB sensors can now sense the angle of the received signal \cite{dwm1000, dwm3000}.
Unfortunately, the entire class of the MDS-based network localization algorithms can not take advantage of this new-found capability of the mobile nodes.
Intuitively the angle information of the peer nodes can serve as additional constraints of the network topology leading to faster convergence to the location estimation.
Moreover, the range and angles are estimated simultaneously per exchange of signals between nodes without incurring additional measurement latency.
This convenient information is wasted as the MDS objective function can not jointly optimize on the Euclidean distance and angle plane.
\todo{We develop a new network localization algorithm with a redefined objective function to include range and angle together for optimization to bridge this gap.}
\new

The scalability of a network localization solution depends on several practical factors beyond the correctness of the theoretical formulations.
The level of dependency on the infrastructure is the most crucial of them.
In addition, a dynamic topology of mobile nodes requires short update latency of location estimations.
Like any peer-to-peer localization system, \name requires at least four anchor nodes for the unambiguous global 3D location of the nodes.
However, the relative locations of the entire topology are correct with no assumption on the anchors.
We enable our algorithms to incorporate any number of available anchor nodes and other infrastructural opportunities to create virtual anchors that enhance overall localization accuracy.
In the proposed system, the joint range and angle-based optimization reduce the measurement and initial topology estimation latency, then a spanning-tree-based optimal edge selection expedites updates on locations after initial estimation, and finally, a graph rigidity-based solution makes the estimation robust to local occlusions or poor peer-to-peer connectivity.
\new

\begin{figure}
    \centering
    \includegraphics[width=1.6in]{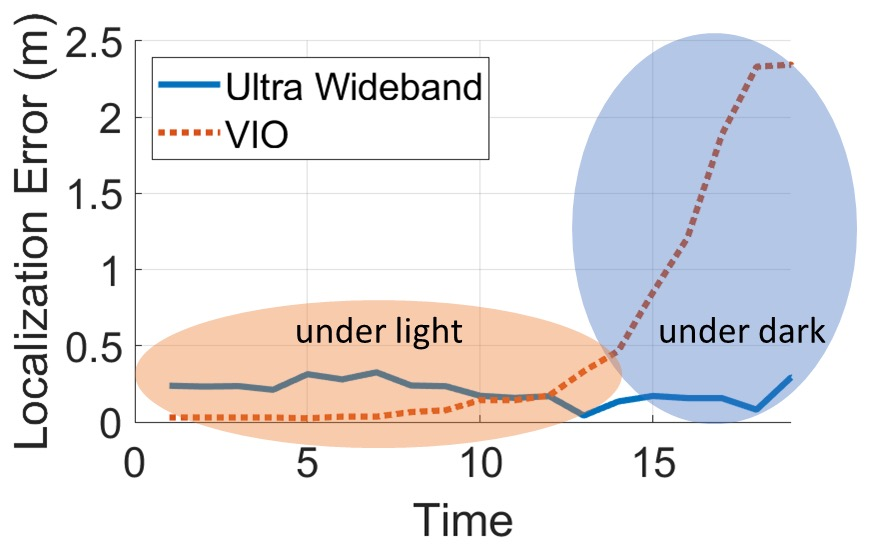}
    \includegraphics[width=1.6in]{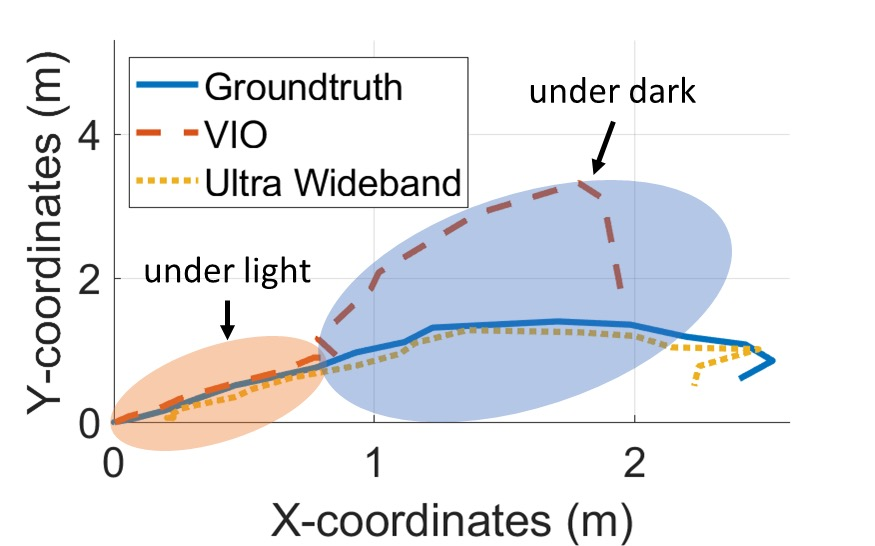}
    \caption{Localization tests for VIO (Visual Inertial Odometry) and UWB (Ultra-Wideband) in varying lighting conditions. (a) The result shows that the VIO system experiences drift and decreased accuracy in darker environments. (b) The result shows the path trajectories during the experiment.}
    \label{fig:feasibility_dark}
\end{figure}

A class of solutions for network localization resort to multimodal data to improve accuracy and reduce the update latency of the system.
While effective in a small number of nodes and within restricted environmental conditions, the multimodality assumption limits the scalability of the system.
It is infeasible to maintain homogeneous data quality with thousands of nodes spread across a large geographical region.
For instance, some recent papers \cite{miller2022cappella} use Visual Inertial Odometry (VIO) - a camera-based solution to track orientation - to improve localization accuracy.
As shown in Figure \ref{fig:feasibility_dark}, in an experiment with VIO and UWB localization the accuracy falters with varying lighting conditions.
We argue that an unimodal solution is ideal for large networks in terms of consistency and ease of practical deployment and maintenance.
It makes our solution equally applicable to the networks of resource-constrained low-power nodes.
The core localization algorithm is, however, applicable to any modality that can measure the peer-to-peer range and relative angles, and naturally, the location accuracy is defined by the accuracy of the measurements.
For instantiating the algorithm in a prototype and realistic large-scale simulations, we consider off-the-shelf Ultra-WideBand (UWB) wireless sensing nodes.
\new

This paper strives to improve the accuracy of 3D localization of UWB-enabled nodes on a large single-modality network as shown in Figure \ref{fig:graphs_comparison_example}.
To this end, we have made the following three specific contributions at the current stage of this project:
\squishlist
\item Developed a novel 3D network localization algorithm to jointly incorporate range and angle in topology estimation. It leads to a fast localization algorithm with 75\% latency improvement for localization of a 30,000-node network spanning several kilometers with $15$ static anchors spread across $14km^2$. The median accuracy of location is $10.2$ meters.
\vspace{-0.1in}
\item Developed supporting algorithm for estimating the optimal spanning tree of the network for robust localization with occlusions and limited field of view of sensor nodes.
The algorithm also includes a decomposition technique a non-rigid graph into smaller rigid graphs, for both range and angle constraints.
\vspace{-0.1in}
\item Implemented a working prototype with UWB-enabled nodes running the proposed algorithm. We used real-world UWB measurement traces to evaluate system performance in large network models.
\squishend

\section{System Design}
\vspace{-0.1in}
Our system, \name, uses pair-wise Ultra-Wideband (UWB) RF measurements, specifically range and angle-of-arrival information, to determine the precise 3D positions of interconnected nodes in relation to one another. The system is designed using four key components:
\new
\textbf{Joint Range-Angle Localization:} We develop an optimization method that uses range and angle measurements as constraints to build a network of localized nodes. Incorporating angles ensures an accurate location estimation with fewer number of edges compared to a range-only method, effectively reducing the latency up to a $4.2\times$.

\textbf{Reference Frame Transformation:} Each node's angle measurement is a composite value of both the relative orientation (determined by the antenna angle with respect to the peer's antenna) and the angle-of-arrival. \name efficiently decomposes this combined measurement to transform all node axis to a common frame of reference.

\textbf{Scaling to Large Networks:} We develop an edge selection strategy that ensures the formulation of the optimal spanning tree, facilitating the interconnection of a large topologies. Large networks often contain flexible or unconstrained edges, which can result in isolated but structurally rigid subgraphs. We aim to identify and resolve these subgraphs separately.

\textbf{Opportunistic Integration with Infrastructure:} \name can leverages any pre-existing infrastructure, incorporating it as anchor points to refine localization accuracy. We further extend our system's capability by introducing the concept of virtual anchors. This mechanism temporarily designates specific mobile nodes as anchors and adjusting edge weights, disseminating the accuracy throughout the network.
Below, we discuss the individual algorithmic contributions in detail.

\subsection{Joint Range-Angle Localization}
\label{subsection:localization_optimization}
\vspace{-0.07in}

We formulate the localization problem as an optimization problem which aims to minimize the difference between measured pairwise distances and the distances corresponding to the estimated coordinates.
There is a plethora of work done in netowrk localization using range measurements only. SMACOF \cite{de2009multidimensional, hamaoui2018non}, non-metric MDS etc \cite{pei2009anchor, beck2014real, sottile2010distributed}. Some of recent works also account for missing edges \cite{shang2004localization, shang2004improved} and NLOS cases \cite{chen2011non, venkatraman2004novel, di2017calibration}. However, none of the works have jointly incorporated angle-of-arrival information in the network.
\new


\textbf{Problem formulation:} Consider a topology of N nodes (mobile devices), with unknown locations $X = [(x_{1}, y_{1}, z_{1}), \cdots (x_{N}, y_{N}, z_{N})]$. Suppose the nodes measure the range between each other, $\hat{r_{ij}}$ where $(i,j)$ is the pair nodes $i$ and $j$ in the topology. We aim to solve for $X$ while minimizing the cost function:

\begin{equation}
    \label{eq:}
    \begin{aligned}
        \underset{X} {\text{min}} \sum_{i, j} w_{ij} (\hat{r_{ij}} - r_{ij}(X))^2
    \end{aligned}
\end{equation}

where, $w_{ij}$ is the weight assigned to edge between nodes $i$ and $j$ and $r_{ij}(X)$ is the euclidean distance between them given by
\begin{equation}
    r_{ij}(X) = \sqrt{(x_{i} - x_{j})^2 + (y_{i} - y_{j})^2 + (z_{i} - z_{j})^2}
    \label{eq:distance_constraint}
\end{equation}
We use $w_{ij} = 0$ for missing edges.
\newline


\textbf{Adding angles in the topology:} Incorporating angles into the network topology offers the benefit of reducing the necessary number of edges, which subsequently decreases latency. This approach, however, is not as straightforward due to the highly non-convex and discontinuous nature of angular loss functions when simply added with the range-based loss function.

Existing works \cite{melnykov2012initializing, biernacki2003choosing, qi2003wireless} which use L1 and L2 losses directly on this loss function show that the objective function has many local minima.

We can compute the angular loss using the function $f(X)$, expressed as:
\begin{equation}
    \label{eq:angle_constraint}
        f(X) = \Big[ \hat{\theta_{ij}} - \text{arctan}\Big[\frac{y_{i} - y_{j}}{x_{i} - x_{j}}\Big] \Big] ^2
\end{equation}

This function encapsulates the difference between the measured angles, denoted by $\hat{\theta_{ij}}$, and the angles calculated from the estimated coordinates.
The non-convex nature of $f(X)$ occurs due to the arctangent and the least square operation on angles.
This results in multiple local minimums, making it highly-prone to generating inaccurate topologies.
The primary contributor to this issue is the restrictive interval $[-\pi/2, \pi/2]$ that the arctangent function operates in, failing to account for points in the left quadrants of the plane, thereby resulting in the same angles for coordinates $(x, y)$ and $(-x, -y)$.
To address this, we first use the 2-argument arctangent, a variant of the arctangent function that considers both x and y inputs and adjusts for the signs, hence returning angles within the inteval $[-\pi, \pi]$.

Unfortunately, the transformation is not enough, as gradient consists of non-differential making it prone to getting ensnared in local minimums.
To overcome this, we apply another transformation to the loss function. We take the negative cosine of the angles, creating a smoother, continuous, and differentiable function restricted within the $[0, 1]$ range.
This transformation of the new loss function, defined as:
\begin{equation}
    \label{eq:orientation_loss}
        f(X) = 1 - cos ( \hat{\theta_{ij}} - arctan2(y_{i} - y_{j}, x_{i} - x_{j} ) )
\end{equation}
Finally, we combine the range and angular loss functions to efficiently integrate angles in the network topology optimization formulating a joint optimization problem.


\begin{equation}
    \label{eq:}
    \begin{aligned}
        \underset{X} {\text{min}} \Big[ \frac{\sum_{i, j} w^{r}_{ij} (\hat{r_{ij}} - r_{ij}(X))^2}{\sum_{i, j} \hat{r_{ij}}} + 
        \sum_{i, j} w^{\theta}_{ij} (1 - cos(\hat{\theta_{ij}} - \theta_{ij}(X))) \Big]
    \end{aligned}
\end{equation}
where, $\theta_{ij}(X) = arctan2(y_{i} - y_{j}, x_{i} - x_{j})$. Note that we scale the range loss function with $(\sum_{i} \sum_{j} \hat{r_{ij}})$. The range loss can be much higher compared to the angular loss, which are in the range $[0, 1]$, and can dominate the overall gradient/overshadow the angular loss. Hence, we normalize the range loss function before adding the angular loss part.
\new

\begin{figure}[tb]
    \centering
    \includegraphics[width=3.1in]{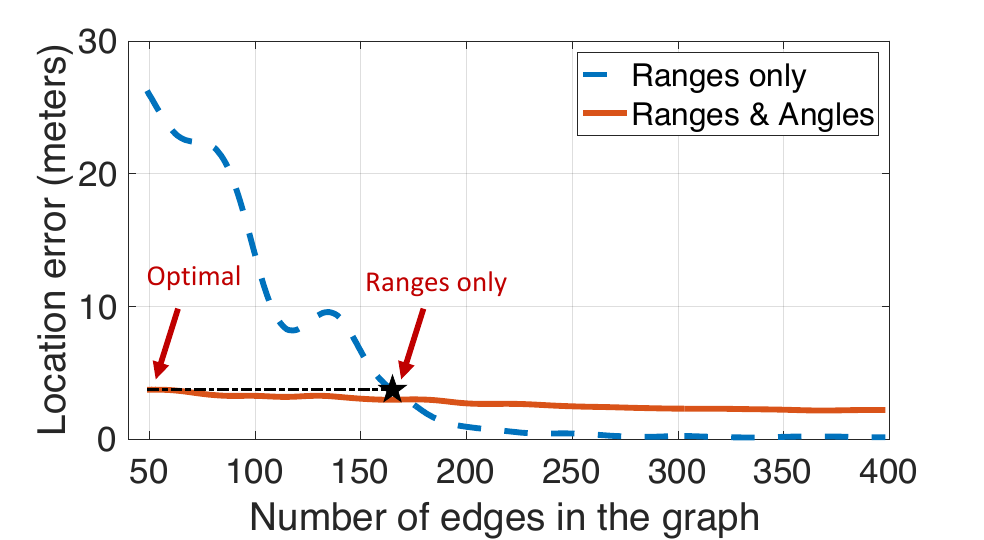}
    \caption{Comparative analysis of constraints: Incorporating angles markedly reduces the number of edges required to attain the same level of accuracy as the "ranges only" approach}
    \label{fig:iterative_adding_edges}
\end{figure}

\begin{figure}[htb]
    \centering
    \vspace{-0.1in}
    \includegraphics[width=3.1in]{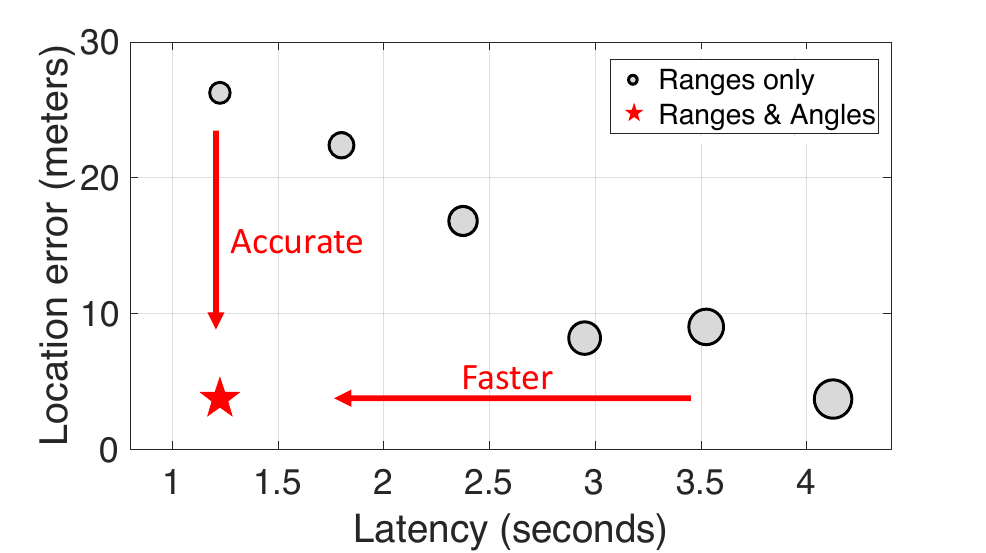}
    \vspace{-0.1in}
    \caption{Using angles significantly improves the latency of the system while maintaining the accuracy.}
    \vspace{-0.1in}
    \label{fig:iterative_adding_edges_2}
\end{figure}

We conducted a simulation analysis to assess the benefit of integrating angles with ranges in topology constraints.
Through the simulation involving a 50-node topology within a $100m\times100m$ region, we see the benefits of this integration.
In the simulation we progressively add edges to the graph and run the optimization it using two distinct constraints: ranges only, and a combination of ranges and angles.
The results, as shown in Figure \ref{fig:iterative_adding_edges}, reveal that utilizing only ranges necessitated more than triple the number of edges to achieve a level of accuracy comparable to that of our method of combining both ranges and angles. Figure \ref{fig:iterative_adding_edges_2} shows a $4.5\times$ improvement in latency over a range-only methos with no compromise in accuracy.
\new

\begin{figure}[tb]
    \centering
    \includegraphics[width=2.9in]{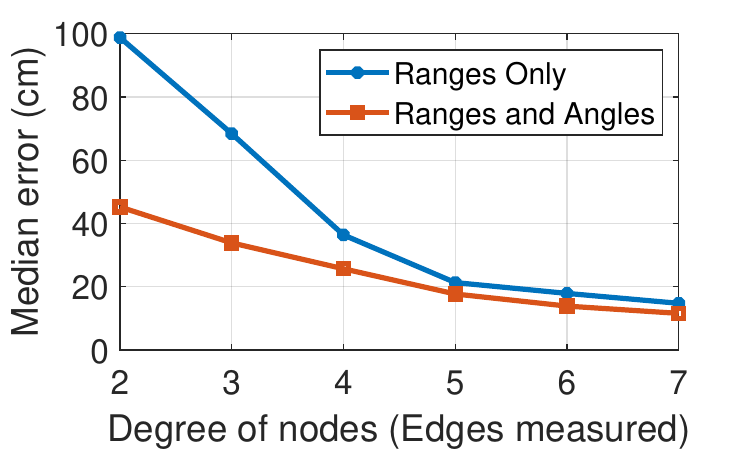}
    \caption{Median localization error for varying number of edge measurements per node.}
    \label{fig:error_vs_degree_of_node}
\end{figure}




\subsection{Scaling to Large networks}
\vspace{-0.07in}
In this section, we detail how \name can localize large node networks efficiently. To adapt to city-scale networks, we address four significant challenges linked with large networks.
$\bullet$ \textit{Optimal Edge Selection}: Identifying the optimal edges to sample in the graph topology is vital to form a unique, rigid structure, while optimizing latency and accuracy.
$\bullet$ \textit{Handling Erroneous Angles}: While in theory all measured edges should have angle data, the limited FOV of commercial antenna arrays can sometimes result in inaccurate angle information. Correcting these errors is vital as they can adversely affect the graph's rigidity constraints.
$\bullet$ \textit{Rigidity Decomposition}: Large networks frequently consist of flexible or unconstrained edges which leads to smaller but rigid subgraphs. We need to identify and solve these subgraphs individually.
Next, we will elaborate on how \name mitigates these challenges to scale to larger networks, introducing techniques that could reduce \name's latency by up to $4.2 \times$.

\subsubsection{Optimal Edge Selection}
During any measurement time iteration, \name compiles a list of optimal edges that are to be measured within the graph. These edges are chosen based on three criteria -
(1) They form a rigid and unique topology, eliminating flexibility or ambiguity.
(2) They are minimal in number to decrease overall latency.
(3) They are feasible, meaning the edges fall within the necessary range and angle FoV.
\new
In a topology comprising n nodes, a maximum of $n(n-1)$ edges are potentially available. However, acquiring data for each possible edge isn't viable as it is time-consuming and the increases exponentially as $n$ increases. Hence, rather than overconstraining the topology, we can create the topology with significantly fewer edges.
According to Laman's condition \cite{laman1970graphs}, for a system that doesn't utilize angles, the minimum number of edges equals to $2n-3$ and $3n-4$ for 2D and 3D setups respectively.
Interestingly, our approach necessitates only a \textit{single} edge per node to fully constrain it, amounting to only $n-1$ edges in total. This efficiency stems from the fact that a single edge encapsulates three constraints - range, azimuth angle, and elevation angle.
Moreover, any random $2n-3$ edges may not always be enough, as some edges could be redundant. Rather we need \textit{well-distributed} $2n-3$ edges to make the topology accurate.
This essentially simplifies our problem, rendering it closely analogous to the \textit{Minimum Spanning Tree problem} found in graph theory \cite{graham1985history}.
\new

\textbf{Minimum Spanning Tree:} 
The Minimum Spanning Tree (MST) is a subset of the graph's edges connecting all nodes with the least total edge weight. In our system, we utilize Kruskal's algorithm, a greedy algorithm that arranges the graph's edges in increasing order of their weights and adds edges to the MST as long as they do not form a cycle, thus guaranteeing the minimum combined edge weights.
\new

\textbf{Edge Weight Calculation:} To minimize the localization error in our MST, we have devised a sophisticated weight assignment algorithm that advances beyond existing noise distribution-based methods \cite{di2017calibration}.
This algorithm considers not only range but also the availability of angles to more accurately determine the \textit{circle of uncertainty} — a metric that encapsulates the potential location uncertainty induced by each edge.
Accordingly, the weight of the edge, denoted as \( w \), is assigned.
The radius of this circle, denoted as \( r \), is formulated by integrating factors such as the edge range value, the angle FOV binary factor (\( f_{\text{$\theta$}} \)) capturing the antenna-array's field-of-view, and the line-of-sight noise distribution (\( \sigma_{\text{los}} \)).
Thus, the edge weight is defined as \( w = \pi r^2 f_{\text{$\theta$}} \sigma_{\text{los}} \). We describe the process in Algorithm \ref{algo:MST}.
\new

\begin{algorithm}[t]
    \caption{Optimal Edge Selection Algorithm}
    \label{algo:MST}
    \hspace*{\algorithmicindent} \textbf{Input:} \( G(V, E) \), list(r, \(f_{\text{$\theta$}}\), \(\sigma_{\text{los}}\)) \\
    \hspace*{\algorithmicindent} \textbf{Output:} Graph \( M(V_{MST}, E_{MST}) \)
    \begin{algorithmic}[1]
        \Procedure{}{}
            \For {each \( E \) in \(G\)}
                \State Retrieve r, \(f_{\text{$\theta$}}\), and \(\sigma_{\text{los}}\) for the current \( E \)
                \State Assign weight \(w(E) = \pi r^2 f_{\text{$\theta$}} \sigma_{\text{los}}\)
            \EndFor
            \State \( G_{sorted}(V, E) \) = Sort(\( G(V, E) \), \(w(E)\))
            \State Initialize \( M(V_{MST}, E_{MST}) \) as empty
            \For {each \( V,E \) in \(G_{sorted}\)}
                \If {adding \( V,E \) to \(M\) does not form a cycle}
                    \State Add \( V,E \) to \(M\)
                \EndIf
            \EndFor
            \State Return \( M(V_{MST}, E_{MST}) \)
        \EndProcedure
    \end{algorithmic}
\end{algorithm}

\begin{figure}[tb]
    \centering
    \includegraphics[width=2.2in]{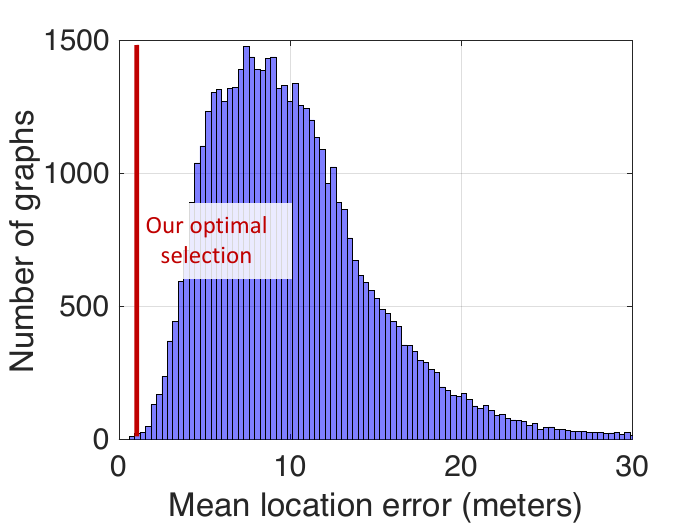}
    \vspace{-0.1in}
    \caption{Histogram of localization errors for all spanning-trees.}
    \vspace{-0.1in}
\label{fig:optimal_selection_spanning_tree_histogram}
\end{figure}

It should be noted that in the proposed algorithm, edges with shorter distances inherently have a smaller circle of uncertainty.
Moreover, edges with available angle information are favored above those restricted to range data only.
Furthermore, Figure \ref{fig:optimal_selection_spanning_tree_histogram} present a comparison of the localization error between our optimal spanning tree and all other possible spanning trees, showing that our algorithm's resulting graph ranks within the top 1\% of all spanning trees.

\begin{figure}[bt]
\centering
\hspace{0.1in}
\subfloat[\small{}]{\includegraphics[width=0.8in]{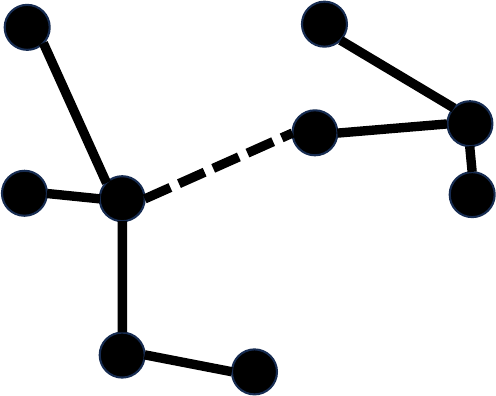}\label{subfig:rigid_topology1}}
\hspace{0.2in}
\hskip1ex
\subfloat[\small{}]{\includegraphics[width=0.8in]{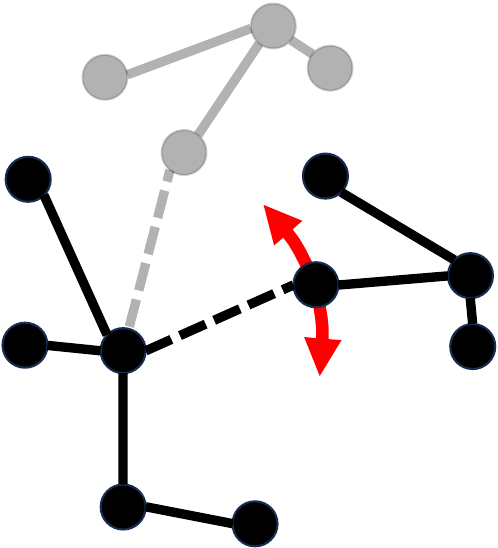}\label{subfig:rigid_topology2}}
\hspace{0.2in}
\hskip1ex
\subfloat[\small{}]{\includegraphics[width=0.8in]{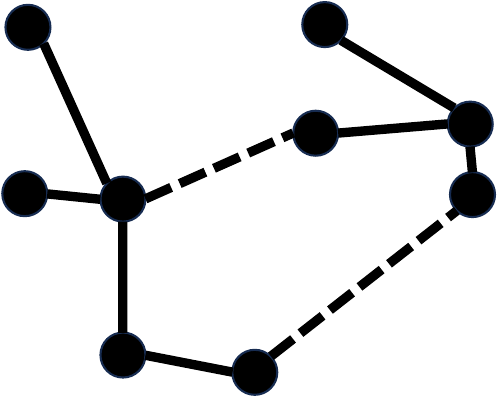}\label{subfig:rigid_topology3}}
\hspace{0.2in}
\hskip1ex
\caption{Different spanning trees representing rigid and non-rigid graphs. Solid lines indicate both range+angle edge, dashed lines indicate range only edge. (a) A connected but non-rigid graph due to missing angle information in an edge. (b) The subgraph is free to rotate. (c) Adding a range measurement makes the graph rigid.}
\label{fig:rigid_graph}
\end{figure}

\subsubsection{Handling Erroneous Angles}
\begin{figure}[tb]
    \centering
    \includegraphics[width=2.6in]{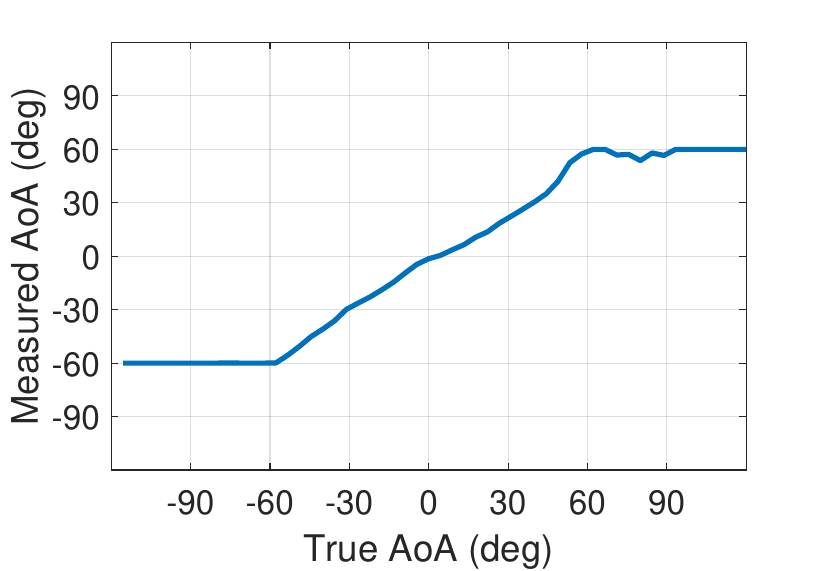}
    \vspace{-0.1in}
    \caption{Reported angles by an off-the-shelf UWB sensor.}
    \vspace{-0.1in}
    \label{fig:fov_feasibility}
\end{figure}

In practical settings, nodes may not consistently measure each other's ranges or angles due to limited sensing range or angular FOV.
This is due to the limited FOV of antenna arrays present in commonly used UWB modules like Decawave\cite{dwm1000, dwm3000} and NXP\cite{nxp_uwb}.
These modules typically restrict the angle-of-arrival field of view (FOV) to between $90^\circ$ and $120^\circ$ to maintain a respectable accuracy in measured AoA, primarily because the AoA estimation algorithm, which depends on the phase difference of the incoming signal, can result in significant errors when the AoA approaches the broadside of the antenna.
Figure \ref{fig:fov_feasibility} shows the reported angles by an off-the-shelf UWB sensor \cite{nxp_uwb}. 
To address this, \name tracks the rotation of each node using its inertial sensors. Utilizing this rotation data we can determine if any two nodes are within each other's FOV and then flag the angles as valid or erroneous accordingly.

\subsubsection{Rigidity Decomposition}
While previous sections operate under the presumption of network topology rigidity, ensuring a unique realization of topology with given constraints, this isn't always the case.
Notably, the optimal spanning tree derived earlier doesn't inherently ensure rigidity.
Even fully connected graphs can remain non-rigid due to absent constraints, as illustrated in Figure \ref{fig:rigid_graph} where spanning trees lack rigidity due to missing angle constraints.
To address this, we develop a rigidity based graph decomposition technique that segments the graph into smaller \textit{rigid subgraphs}.
These subgraphs can then be solved with refined edge weights ensuring that the topology is both rigid and uniquely realizable.
\new



\textbf{Quantifying Rigidity:} Rigidity of a topology determines whether it is possible to uniquely determine the location of all nodes in the topology with respect to other node.
This leaves us with some trivial transformations of topology like translations and rotations in the space. 
The rigidity can be quantified using the degree of freedom (DoF) of the graph.
Each node has three DoFs, representing movements in the x, y, and z directions.
Thus, a graph with \( n \) nodes inherently has \( 3n \) DoFs.
As constraints are imposed, the available DoFs decrease.
A graph is deemed rigid if its DoFs are \( \leq 3 \), with these residual DoFs accounting for the whole-graph translational motions.
\new

\textbf{Eigenanalysis of Rigidity Matrix:} To identify independent and unconstrained motions that the topology can perform without violating the constraints, we perform the eigenvector analysis of the topology.
To capture the constraints in a compact matrix form, we first introduce the rigidity matrix, \(R\).
Every row of \(R\) denotes a unique constraint equation, which could be a distance or angle constraint between nodes.
Each column of \(R\) corresponds to the \(x\), \(y\) and \(z\) coordinates for each node in the graph. Thus, for a graph having n nodes, the matrix expands to have \(3n\) columns.

Each entry \(r_{ij}\) in this matrix is determined by the partial derivatives of the constraint equations \ref{eq:distance_constraint} and \ref{eq:angle_constraint}, as given by:


\[
r_{ij} = 
\begin{cases} 
  \frac{\partial d_{ij}}{\partial x_m} & \text{if edge } ij \text{ is a distance constraint } \\
  \frac{\partial \theta_{ij}}{\partial y_m} & \text{if edge } ij \text{ is an angle constraint} \\
  0 & \text{otherwise}
\end{cases}
\]


\[
R = 
\begin{bmatrix}
\frac{\partial d_{1,2}}{\partial x_1} & \frac{\partial d_{1,2}}{\partial y_1} & \cdots & \frac{\partial d_{1,2}}{\partial x_n} & \frac{\partial d_{1,2}}{\partial y_n} \\
\frac{\partial d_{2,3}}{\partial x_1} & \frac{\partial d_{2,3}}{\partial y_1} & \cdots & \frac{\partial d_{2,3}}{\partial x_n} & \frac{\partial d_{2,3}}{\partial y_n} \\
\vdots & \vdots & \ddots & \vdots & \vdots \\
\frac{\partial \theta_{1,2}}{\partial x_1} & \frac{\partial \theta_{1,2}}{\partial y_1} & \cdots & \frac{\partial \theta_{1,2}}{\partial x_n} & \frac{\partial \theta_{1,2}}{\partial y_n} \\
\frac{\partial \theta_{2,3}}{\partial x_1} & \frac{\partial \theta_{2,3}}{\partial y_1} & \cdots & \frac{\partial \theta_{2,3}}{\partial x_n} & \frac{\partial \theta_{2,3}}{\partial y_n} \\
\vdots & \vdots & \ddots & \vdots & \vdots \\
\end{bmatrix}
\]

\begin{figure*}[tb]
    \centering
    \includegraphics[width=2.2in]{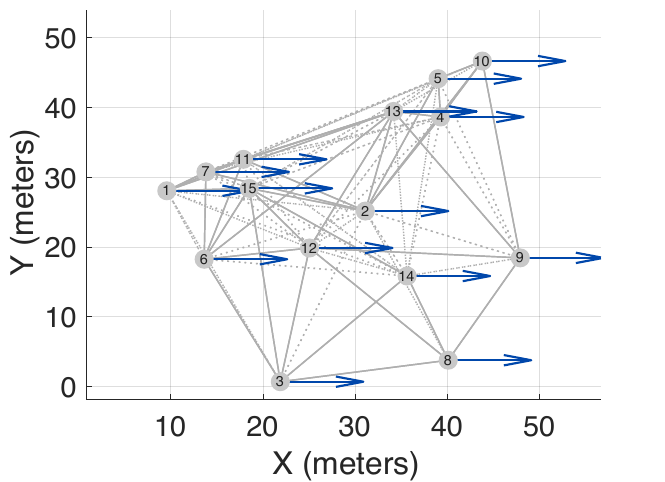}
    \includegraphics[width=2.2in]{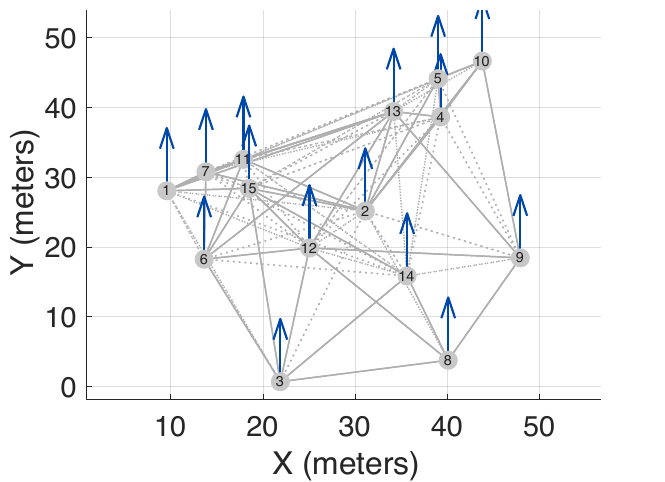}
    \includegraphics[width=2.2in]{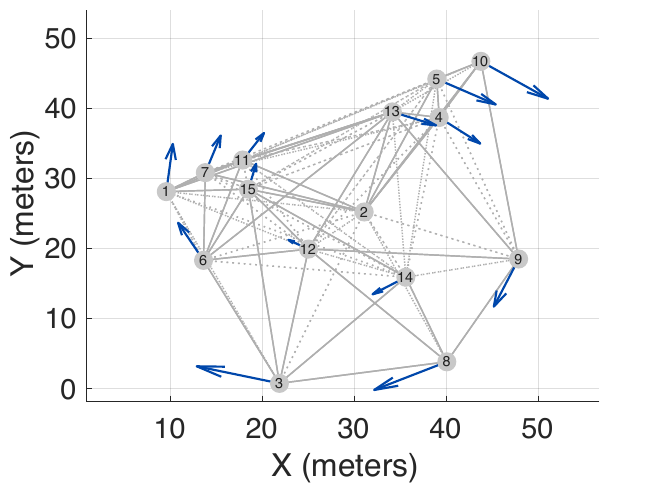}
    \caption{{The displacements corresponding to zero eigenvalues represent the translational and rotational motions that the nodes can undergo without violating any constraint.}}
    \label{fig:eigen_value_motions}
\end{figure*}

Let \( \lambda_1, \lambda_2, \dots, \lambda_m \) be the eigenvalues of the matrix \( M = RR^T \), where $M$ is the symmetric and positive semi-definite matrix of $R$.
The topology is uniquely rigid if the number of zero eigenvalues of $M$ is equal to $3n-3$.
The count of zero eigenvalues, or the degree of freedom, can be represented as:

\begin{equation}
DoF = \sum_{i=1}^{m} \delta(\lambda_i)
\end{equation}
where,
\[
\delta(\lambda) = 
\begin{cases} 
  1 & \text{if } |\lambda| < \epsilon \\
  0 & \text{otherwise}
\end{cases}
\]
Here, \( \epsilon \) is a small positive number close to zero, chosen to account for numerical inaccuracies (e.g., due to floating-point representation limits in computers). If \( |\lambda_i| < \epsilon \), it is considered a zero eigenvalue.
Figure \ref{fig:eigen_value_motions} illustrates the eigenvectors corresponding to the three zero eigenvalues \( \lambda_1, \lambda_2, \text{ and } \lambda_3 \) showing the motion of each node.
\new

\textbf{Constructing Subgraphs:} By examining the zero eigenvalues, we gain insights into the \textit{displacement vectors} of each node.
For a fully rigid topology, these vectors maintain consistent magnitudes and directions across nodes.
In contrast, non-rigid graphs display varied displacements.
Our core intuition is that displacement vectors can be used for decomposition: nodes that have identical displacement vectors in terms of magnitude and direction inherently form a rigid subset.
This means they can only undergo collective motion.
To extract and group these subsets, we first reshape the eigenvector into a 3×n matrix to extract the displacement vectors of each node.
Each row within this matrix signifies the displacement of a node across the three axes.
Next, we identify the unique displacement vectors in this matrix where every unique vector signals the presence of a distinct "rigid subgraph" within the primary graph.
By grouping the nodes corresponding to each unique displacement vector within the matrix, we find the vertices belonging to the associated "rigid subgraph".
\new

\textbf{Critical Edges List:}
Beyond recognizing rigid subgraphs, our approach also pinpoints the edges connecting the separate subgraphs.
We term these edges as 'critical edges', as these edges are crucial for rigidity, serving as the connectors between independent rigid substructures.
For enhanced analysis in subsequent iterations, we also prioritize these critical edges based on their latest computed locations.
This prioritized list of edges underlines those edges that are more likely to come into proximity with one another, thus enhancing the overall scalability and rigidity of the graph.

\subsection{Reference Frame Transformation}
\begin{figure}[htb]
    \centering
    \includegraphics[width=2.9in]{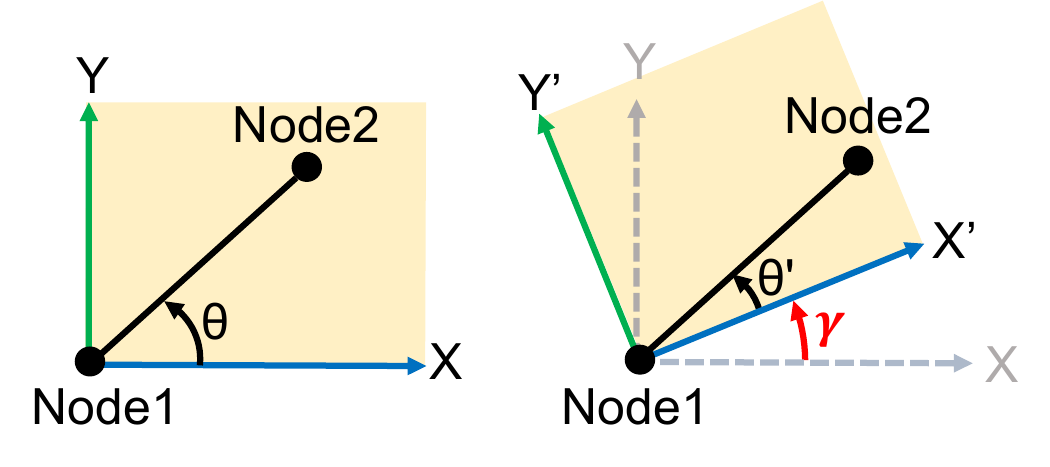}
    \caption{(a) When Node 1 is perfectly aligned with the global frame of reference, it reports that Node 2 is positioned at angle $\theta$. (b) However, when Node 1 is rotated by $\gamma$ relative to the global frame of reference in a 3D space, it reports a distinctly different angle $\theta'$, for the same Node 2.}
    \label{fig:orientation_problem}
\end{figure}

\vspace{-0.07in}
In the above section, we assume that all nodes shared a common frame of reference.
Here, we relax this assumption and address the challenges arising from differing local frames of reference for each node.
Every node measures angles based on two factors: its relative orientation (defined by its antenna angle concerning another node's antenna) and the angle-of-arrival.
The challenge lies in the fact that these angle measurements are in the node's local frame of reference, which moves according to the node's orientation within a broader, global context.
Consider the 2D example in Figure \ref{fig:orientation_problem}.
In scenario (a), Node 1 identifies Node 2 at an angle $\theta$. But in scenario (b), after Node 1 turns by an angle $\gamma$, it sees Node 2 at a different angle $\theta'$.
This change in reported angles is due to Node 1's rotation, even though the positions of the nodes didn't change.
This issue gets complex in 3D, where nodes can rotate in any direction.
Our goal is to deconstruct these angle measurements from orientation offsets and align all nodes to a singular, shared frame of reference.
\new

\subsubsection{Estimating Orientation Offsets}
Every node's orientation in a 3D space is defined by a set of three rotation angles. Specifically, they depict the node's rotations about its X, Y, and Z axes and are recognized as the Roll$(\alpha)$, Pitch$(\beta)$ and Yaw$(\gamma)$ \cite{weisstein2009euler}.
To estimate a node's orientation, we leverage a fundamental observation: When the orientations of nodes are in the same global frame, the measured azimuth angle-of-arrival ($\theta$) and elevation angle-of-arrival ($\phi$) are complementary, implying they obey the condition: $\theta_{ij} - \theta_{ji} = \pi$ and $\phi_{ij} = - \phi_{ji}$
Using this insight, we formulate an optimization problem with the goal to estimate the rotation offsets $(\alpha, \beta, \gamma)$. The objective is to minimize the deviation from the above-mentioned constraints:
\new
\begin{equation}
    \label{eq:orientation_loss}
    \begin{gathered}
        \underset{\alpha, \beta, \gamma} {\text{min}} \sum_{i, j}
        (\hat{\theta}_{ij} - \hat{\theta}_{ji} -\pi) + (\hat{\phi}_{ij} + \hat{\phi}_{ji})
    \end{gathered}
\end{equation}

In this equation, $\hat{\theta}{ij}$ and $\hat{\phi}{ij}$ denote the rotated azimuth and elevation angles after the node has undergone rotation by $\alpha_{i}$, $\beta_{i}$, and $\gamma_{i}$. Solving for this optimization provides us with the required orientation offsets to align all nodes in the system with the global frame of reference. The residual angles, $\hat{\theta}$ and $\hat{\phi}$, thus correspond to the AoAs in the global frame.
\new

Next, we breakdown the steps in optimization problem, detailing how we compute $\hat{\theta}$ and $\hat{\phi}$ through the rotation of local frames.
The rotation in a 3D context can be decomposed into three unique rotations around the x, y, and z basis axes.
\new
\textit{Step1} - Determine the unit vector based on the estimated azimuth and elevation angles:
\[ U = 
\begin{bmatrix}
\hat{x}\\ \hat{y}\\ \hat{z}
\end{bmatrix}
=
\begin{bmatrix}
cos(\phi)*cos(\theta)\\
cos(\phi)*sin(\theta)\\
sin(\phi)
\end{bmatrix}
\]

\textit{Step2} - Apply the rotation matrix \cite{slabaugh1999computing} to adjust the unit vector using the computed roll, pitch, and yaw angles ($\alpha$, $\beta$, and $\gamma$). This results in a new unit vector, $V$, which reflects the node's orientation post adjustment.

\[V = R_{z}(\gamma)R_{y}(\beta)R_{x}(\alpha) U \]

\begin{equation*}
\resizebox{3.2in}{!}{$
V = 
\begin{bmatrix}
\cos{\gamma} & -\sin{\gamma} & 0 \\
\sin{\gamma} & -\cos{\gamma} & 0 \\
0 & 0 & 1 \\
\end{bmatrix}
\begin{bmatrix}
\cos{\beta} & 0 & \sin{\beta}\\
0 & 1 & 0\\
-\sin{\beta} & 0 & \cos{\beta}\\
\end{bmatrix}
\begin{bmatrix}
1 & 0 & 0 \\
0 & \cos{\alpha} & -\sin{\alpha}\\
0 & \sin{\alpha} & -\cos{\alpha}\\
\end{bmatrix}
U
$}
\end{equation*}

\textit{Step3} - Finally, compute the azimuth and elevation angles from the rotated vector V to obtain the modified angle measurements.
\[ 
\begin{bmatrix}
\hat{\theta} \\ \hat{\phi}
\end{bmatrix}
=
\begin{bmatrix}
tan^{-1}(y/z)\\
tan^{-1}(z/\sqrt{x^2 + y^2 + z^2})
\end{bmatrix}
\]
\new

\begin{figure}[htb]
    \vspace{-0.1in}
    \centering
    \includegraphics[width=2.9in]{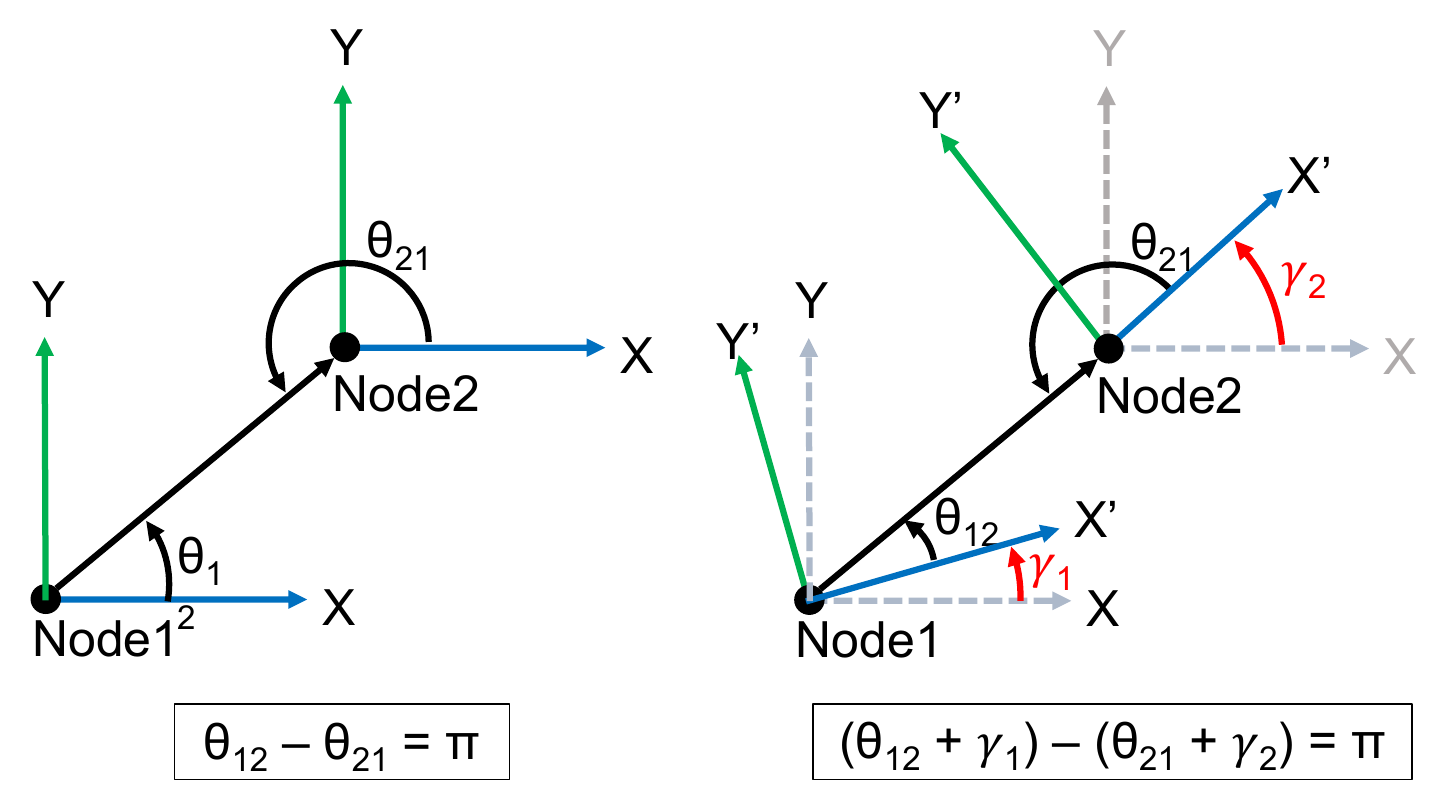}
    \vspace{-0.1in}
    \caption{The relation between AoAs reported by two nodes (a) when nodes are perfectly aligned, and (b) when nodes are oriented with angle $\gamma$ relative to the global frame of reference.}
    \vspace{-0.1in}
    \label{fig:orientation_example}
\end{figure}

If we already have roll and pitch data from another source, the optimization process becomes more straightforward.
Instead of considering all three rotation angles, we only focus on the yaw angle, represented by $\gamma$.
As shown in Figure \ref{fig:orientation_example}, when two nodes are perfectly aligned with the global frame of reference, the difference between their reported AoAs is a constant, $\pi$.
However, if these nodes are rotated by certain yaw angles, their reported AoAs will change exactly equal to their orientation.
\begin{equation}
    \label{eq:orientation_loss}
    \begin{gathered}
        \underset{\gamma} {\text{min}} \sum_{i, j}
        (\pi - (\theta_{ij} + \gamma_{i}) - (\theta_{ji} + \gamma_{j}))
    \end{gathered}
\end{equation}
Using these new offsets, we can now adjust the reported AoA readings to align with the global frame of reference.

The simplification is grounded in the insight that when two nodes are aligned the global frame of reference (implying that the $\gamma$ angle is zero), the AoAs reported by these nodes follow a constraint: $\theta_{ij} - \theta_{ji} = \pi$.
However, when nodes are rotated by arbitrary yaw angles $\gamma$, their reported AoAs adhere to a new relationship that accounts for the node's orientation: $(\theta_{ij} + \gamma_{i}) - (\theta_{ji} + \gamma_{j}) = \pi$, also shown in Figure \ref{fig:orientation_example}. 
By initializing the optimization process with $\gamma_{1} = 0$, all subsequent orientations are determined within the reference frame of the first node, resulting in precise orientation estimates for all nodes.
Subsequently, by subtracting the estimated yaw angle from the reported AoA readings ($\hat{\theta}_{ij} = \theta_{ij} - \gamma_{i}$), we attain AoAs that are relative to the global frame of reference.

\section{Opportunistic Anchor Integration}
\vspace{-0.1in}
\name is specifically designed to not necessitate a static anchor, but it can seamlessly integrate the presence of an existing anchor to refine the location and orientation results and transition from relative to global coordinates.
This capability also offers an edge over range-only systems, which require at least three anchors for global coordinates which is a challenging requirement in highly mobile environments.

We categorize anchors into two types: (1) Static Anchors and (2) Virtual Anchors.

\subsection{Static Anchors}
\vspace{-0.07in}
Static anchors are conventional infrastructure anchors with known locations. These anchors have increased edge weightage, enhancing their likelihood of being selected over mobile volatile edges. The impact of these anchors is further elaborated in the Evaluation Section.

\subsection{Virtual Anchors}
\vspace{-0.07in}
Since many new infrastructure cameras are getting connected to 5G, we take advantage of them to perform accurate localization in the vicity and assign the users which are in the field of view of the camera as \textit{virtual anchors}. A virtual anchor is simply a user which has high confidence score in location obtained by other modality like an infrastructure camera.
\new

\textbf{User registration:}
Leveraging the connectivity of modern infrastructure cameras to 5G, an application can use them for precise localization. Users within the camera's field-of-view are designated as \textit{virtual anchors}. Essentially, a virtual anchor is a user with a high-confidence location sourced from an external modality, like an infrastructure camera.
\new

\begin{figure}[tb]
    \centering
    \includegraphics[width=3.1in]{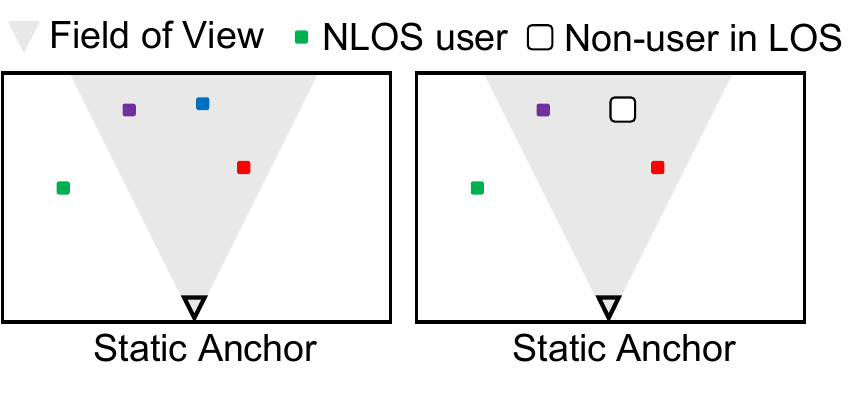}
    \vspace{-0.1in}
    \caption{{Different scenario of users visible to a static camera anchor.}}
    \label{fig:users_registration_depiction1}
\end{figure}

\begin{figure}[bt]
\centering
\subfloat[\small{Time taken}]{\includegraphics[width=1.6in]{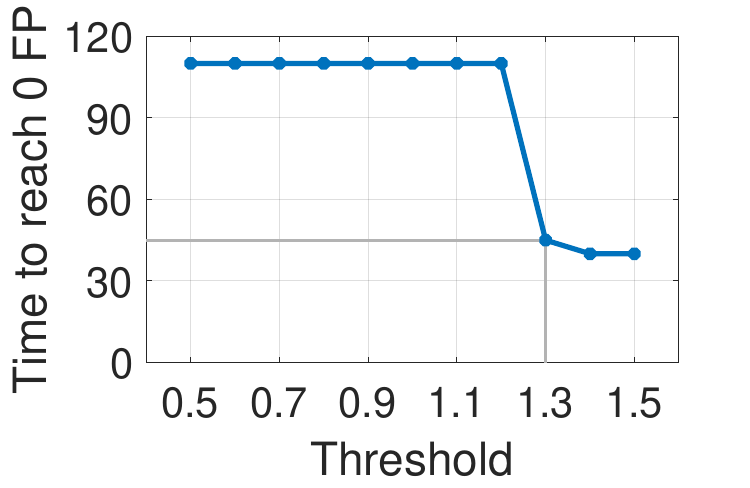}\label{subfig:users_registered_time_vs_threshold}}
\subfloat[\small{Percent of users}]{\includegraphics[width=1.6in]{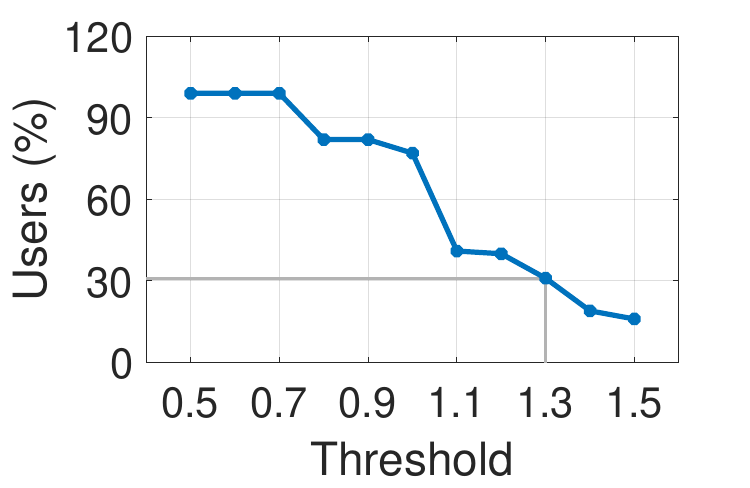}\label{subfig:users_registered_vs_threshold}}
\vspace{-0.1in}
\caption{{User registered for different thresholds of trajectory matching.}}
\vspace{-0.1in}
\label{fig:User_registeration}
\end{figure}

\textbf{Technique:}
Our registration technique uses correlation in user trajectories to incorporate human motion dynamics such as varying walking speeds and stationary periods.
The system benefits from \textit{heading direction}, \textit{cosine similarity of motion} and speed analysis.
In Figure \ref{fig:users_registration_depiction1}, we observe a scenario where some users are equipped with our node, while others are not, all within the camera's Field of View (FoV).
Our primary objective is to accurately register the system users in the camera's FoV.
It's crucial to avoid misclassifications here, as any incorrect virtual anchor can be misdirect our system's optimizer.
To address this, we adopt a stringent approach to trajectory similarity.
Our aim is to ensure zero false positives (FP).
As depicted in Figure \ref{fig:User_registeration}, employing a higher threshold for registration effectively achieves zero FPs.
By analyzing 45 seconds of motion data, we successfully register approximately $30\%$ of visible users, ensuring the reliability of our system.

\section{Implementation}
\label{sec:implementation}
\vspace{-0.1in}
To this end, we implement a prototype of \name and perform experiments in different small and large scale scenarios.

\begin{figure}[htb]
    \centering
    \includegraphics[width=1.48in]{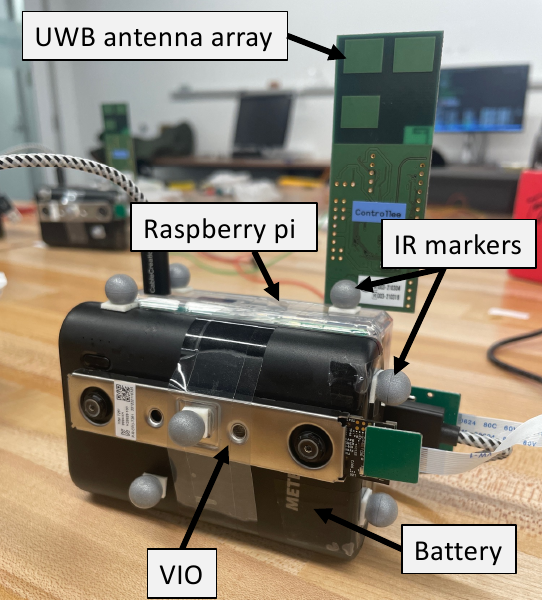}
    \hspace{0.02in}
    \includegraphics[width=1.75in]{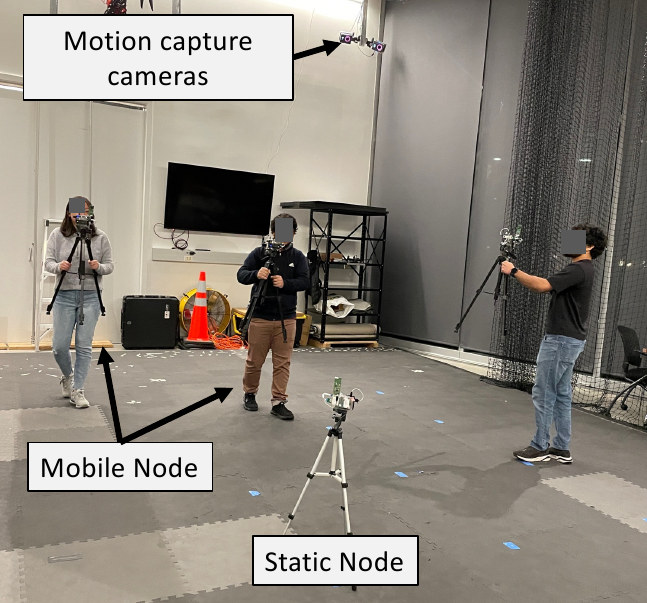}
    \vspace{-0.2in}
    \caption{Our evaluation setup (a) The mobile sensor node is build on a Raspberry Pi, UWB antenna array, Intel Realsense VIO, IR markers for ground truth and a battery pack. (b) One of the test environment with Vicon motion capture.}
    \label{fig:evaluation_setup}
\end{figure}

\textbf{Node prototype:} We developed prototype mobile nodes of \name to evaluate the performance. Each node consists of a Raspberry pi 3 \cite{raspberrypi} for computation, an Intel Realsense T261 \cite{intelrealsense} to collect visual inertial odometry data. We use the NXP SR150 UWB evaluation boards\cite{nxp_uwb}
to collect the range, azimuth and elevation AoA data. We collect the UWB measurements at a rate of 20Hz. The overall measurement rate depends on the number of nodes in topology. All the nodes are time synced using NTP.
\new

\textbf{Software:}
All the prototype nodes are part of a a network and communicate with each other using over a UDP socket. We collect and store the UWB range and AoA data and VIO data from all nodes using a python script. We then process all measurements offline using Matlab. We implement all our algorithms using Matlab's optimization toolbox.
\new

\textbf{Ground truth:} We use the Vicon motion capture setup \cite{viconv8} consisting of 12 Vantage V8 cameras that are mounted on the ceiling to track and record the ground truth locations and orientations of all nodes. The cameras uses IR light to track reflective markers which are attached to the nodes. We use 9 markers per node to reliably track the nodes in all orientations. The measurement accuracy of the setup is under 1mm and the frame rate is 200Hz.
\new

\textbf{Evaluation environments:}
We evaluate the system's localization performance in various types of environments including LOS and NLOS scenarios and varying lighting conditions. We evaluate the system in an indoor setting with upto 6 users walking naturally in a random path. The rooms dimensions of our locations are 6m x 8m and 4m x 20m with various kinds of furniture and equipment in the environment. During data collection, we also have additional users who are not a part of the system, walking in the environment to emulate a crowded and dynamic environment.
\new

\textbf{Large scale simulation analysis:}
We have also implemented a simulation testbed that draws from real-world peer-to-peer measurements, spanning various conditions like range variations, AoA discrepancies, and LOS/NLOS situations. Our dataset comprises of both precise ground truth values and their corresponding noisy ranges and AoA readings.
During all our simulations, we adhere to these real-world measurements. For instance, in NLOS scenarios, nodes select an AoA reading from the NLOS subset that aligns with the particular distance and angle, incorporating contextual noise.
We will open source our new simulation software and the real-world datasets.
\section{Evaluation}
\vspace{-0.1in}
We conduct experiments under different conditions such as line-of-sight (LOS) and non-line-of-sight (NLOS), stationary and moving nodes. We evaluate the impact of varying numbers of edges and number of virtual anchors on localization performance. We have collected a total of 2 hours of real world data with a motion capture setup to extensively evaluate the centimeter level localization accuracy and sub degree level orientation accuracy of our system. We use the closest work \cite{miller2022cappella} which uses UWB+VIO for multi-user localization as our baseline comparison and compare the two in practical scenarios. 
We also perform large city-scale simulation analysis using the real world data to evaluate how does our system scale.
Next, we provide further details of our evaluations.

\begin{figure}[tb]
    \centering
    \includegraphics[width=1.6in]{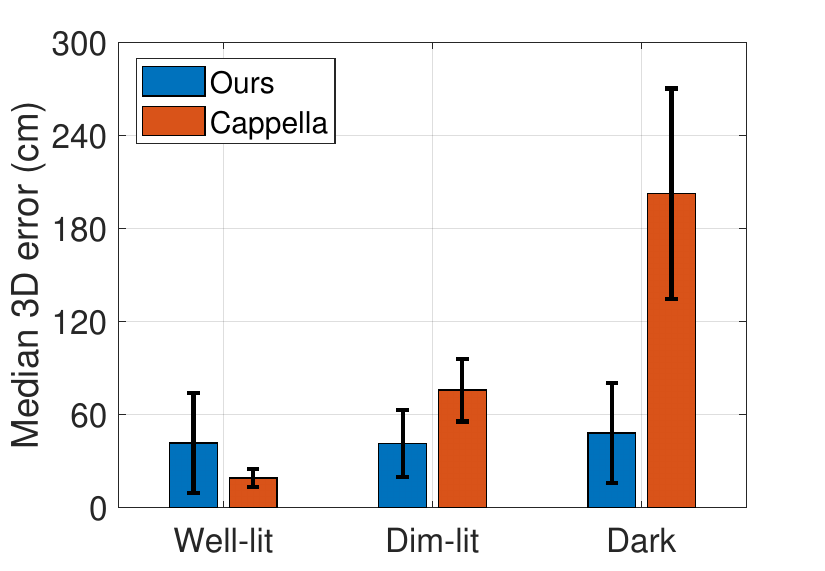}
    \includegraphics[width=1.6in]{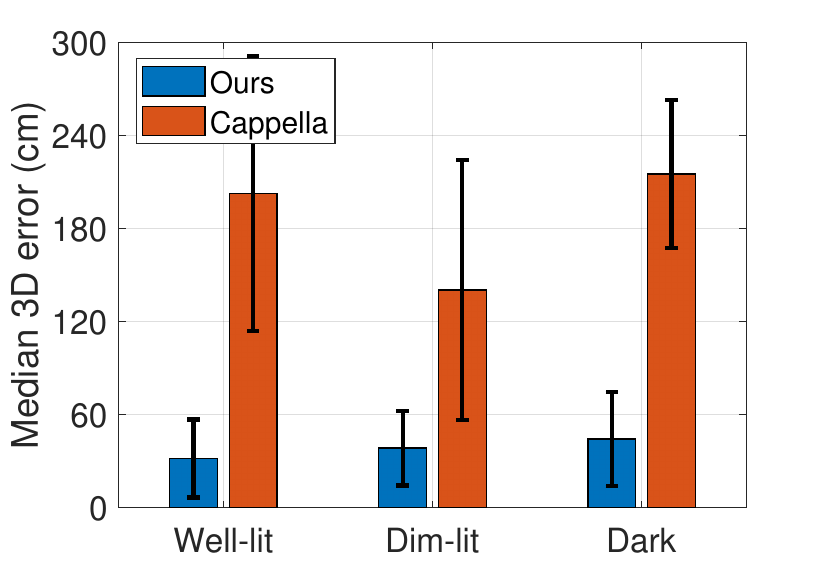}
    \caption{3D Localization performance of \name compared to Cappella \cite{miller2022cappella}, which depends on visual odometry along with UWB. Results shows performance in different lighting condition and in (a) dynamic and (b) static networks.}
    \label{fig:baseline_comparison_lighting}
\end{figure}


\subsection{Localization accuracy}
\vspace{-0.07in}
\textbf{Comparison with Baseline (Camera based system):} Figure \ref{fig:baseline_comparison_lighting}(a) shows the median localization error for \name and the baseline \cite{miller2022cappella}. We compare the errors for varying lighting conditions to show that our system is agnostic to any lighting conditions whereas Cappella's localization errors increase with decrease in illumination. This is mainly due to the over reliance of the baseline system on visual features. Our approach is dependent of RF based peer-to-peer measurements and is not affected by any visual condition. Similarly, we show in Figure \ref{fig:baseline_comparison_lighting}(b) that when nodes are static and the baseline system does not has a tail of trajectory, it fails to estimate 3D location. \name on the other hand does not rely on odometry and can estimate the complete 6DOF location and orientation for all the nodes whether it is static or mobile.
\new

\begin{figure}[tb]
    \centering
    \includegraphics[width=1.6in]{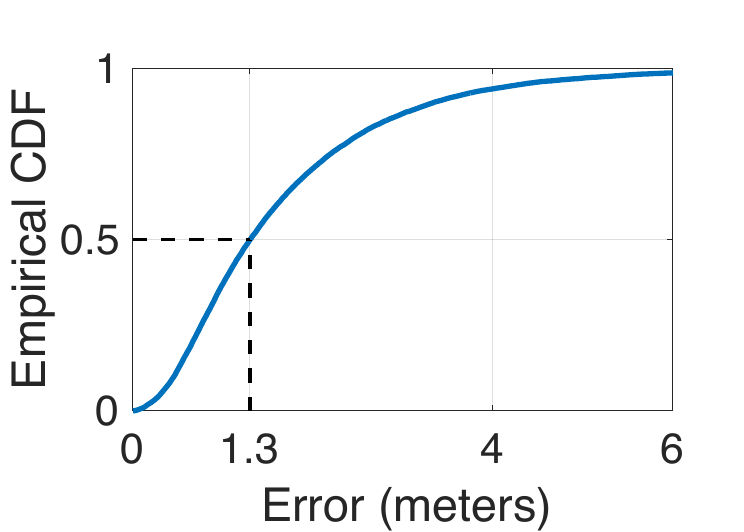}
    \includegraphics[width=1.6in]{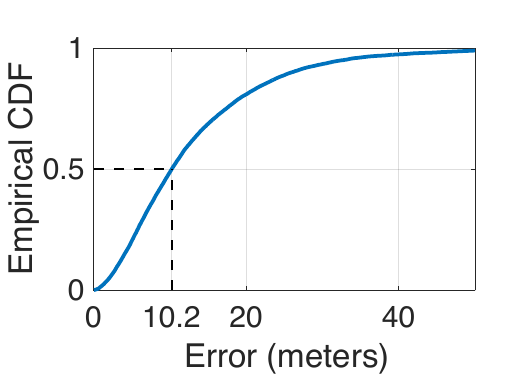}
    \caption{{City-scale analysis: Localization errors of $30,000$ nodes in $3800 \times 3800 \times 200$ meter 3D space. (a) Using $10\%$ of nodes as anchors (b) Using $0.05\%$ of nodes as anchors.}}
    \label{fig:city_scale_simulation}
\end{figure}

\textbf{City-level simulation:}
We evaluate the performance \name in city-scale scenarios by emulating a 3800 x 3800 x 200 meter 3D environment with real world collected data infused in the simulation.
We simulated a total of $30,000$ nodes distributed randomly, as described in section \ref{sec:implementation}.
Figure \ref{fig:city_scale_simulation} shows the localization accuracy of the nodes with different number of infrastructure anchors.
The median localization error is 1.3 meters.
Furthermore, even when we reduce the number of anchors to a mere $0.05\%$ of the total nodes, which is 15 anchors in the city, the localization error remains well within the decimeter range.
\new

\begin{figure}[htb]
    \centering
    \includegraphics[width=2.9in]{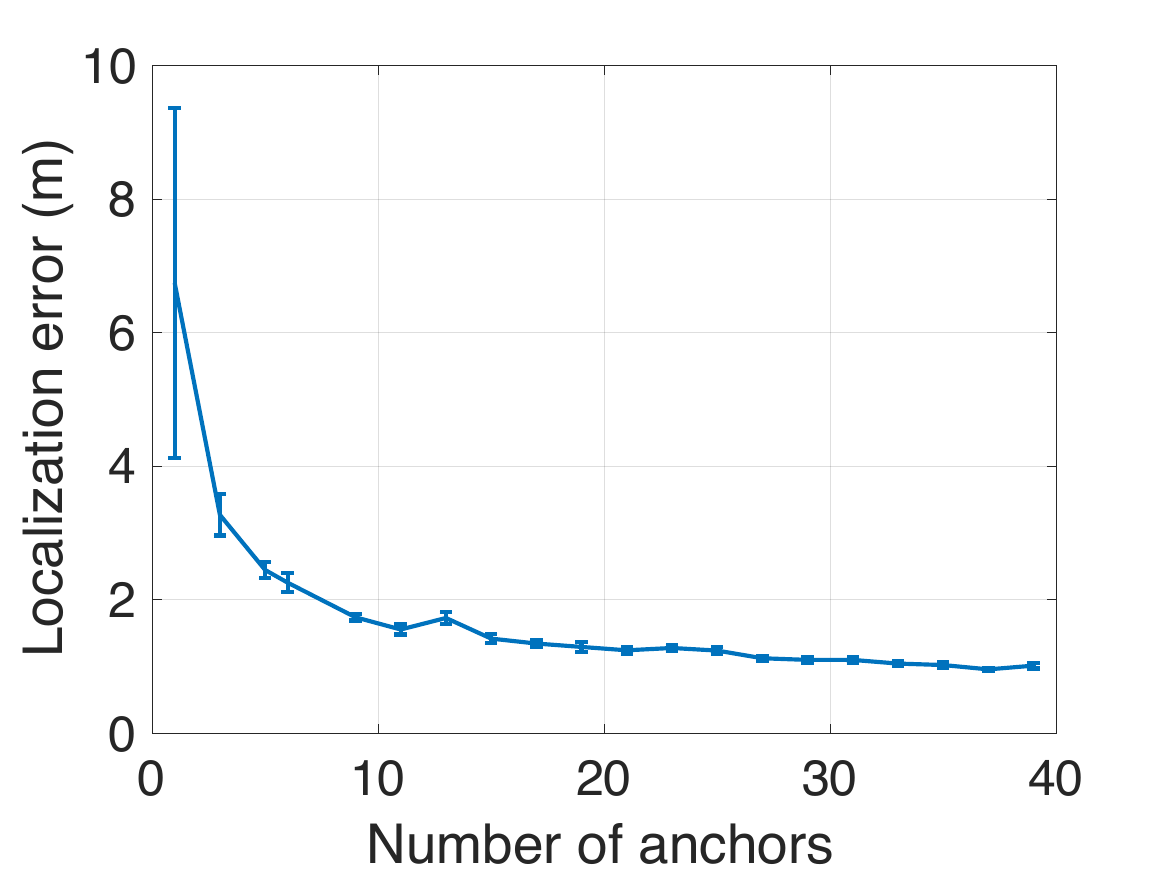}
    \caption{{City-scale analysis: Localization errors of a 1000 node topology in a $200 \times 200 \times 50$ meter 3D space for varying number of static anchors.}}
    \vspace{-0.1in}
    \label{fig:city_scale_simulation_varying_anchors}
\end{figure}

\textbf{Varying anchor density:}
We investigate the impact of node density on the localization performance of \name.
To achieve this, we set up an emulated scenario consisting of 1,000 nodes distributed within a 3D space measuring 200 x 200 x 50 meters.
Figure \ref{fig:city_scale_simulation_varying_anchors} reveals an exponential decay in localization error with the increase in the number of nodes. 
With only 4 nodes, the localization error drops below the 2-meters.
\new

\begin{figure}[htb]
    \vspace{-0.1in}
    \includegraphics[width=1.6in]{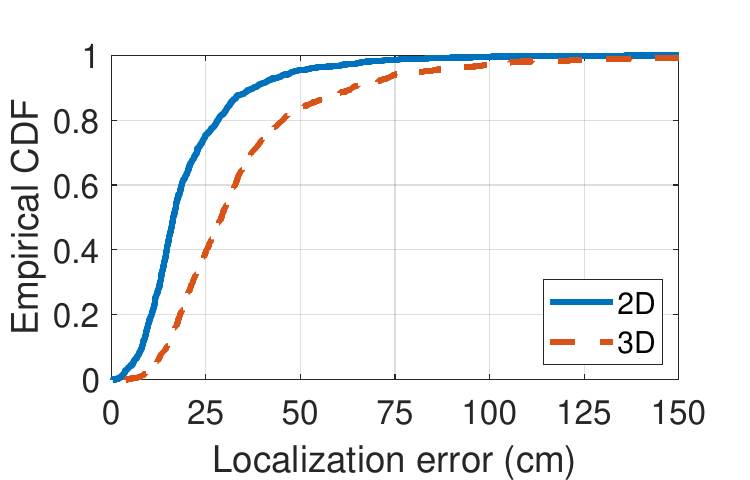}
    \includegraphics[width=1.6in]{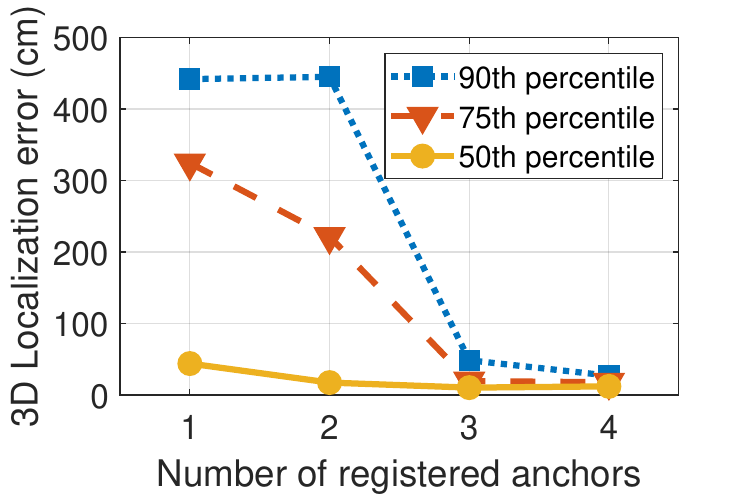}
    \caption{(a) CDF location errors. (b) 3D Localization error for varying number of virtual anchors registered.}
    \label{fig:overall_location}
\end{figure}

\textbf{2D vs 3D accuracy:} We evaluate the 2D and 3D location estimation performance of \name by comparing the euclidean distance errors of each node in the topology from the ground truth. Since our system is infrastructure free and does not require any anchors for localization, it estimates the node locations is a local reference frame. To compute the errors with the ground truth, we rotate and translate the estimated topology into a global reference frame and then evaluate the location errors for each node in the topology. Figure \ref{fig:overall_location}(a) shows the cumulative density function (CDF) of the 3D and 2D localization errors. The median errors for 2D is less compared to the 3D errors. It is due the additional dimension of height which the optimization algorithm solves. This evaluation contains 83 minutes of real world data containing different scenarios of LOS, NLOS, static and mobile nodes. Overall the median errors are $18cm$ and $30cm$ and $90^{th}$ percentile errors are $38cm$ and $65cm$ for 2D and 3D respectively.
\new

\textbf{Impact of virtual anchors:} Figure \ref{fig:overall_location}(b) shows the localization error with varying number of virtual anchors registered by the system. Virtual anchors help is orienting the topology in an absolute coordinate space instead of a local coordinate space. Ideally, we need atleast 3 anchors to remove ambiguities in 3D orientation of the topology. But in our system, since we get angles of the topology along with the location, we could also peg the entire topology using only two anchors.
When we have only one anchor to peg the topology, the main source of error is the rotation of topology. The location errors of the anchor is reduced when we get the virtual anchor. However, due to any error is the angle of the virtual anchor, the entire topology is rotated. Hence the error in case of one anchor is higher. As we register more number of anchors, the accuracy gets better. 
\new

\begin{figure}[htb]
    \includegraphics[width=1.6in]{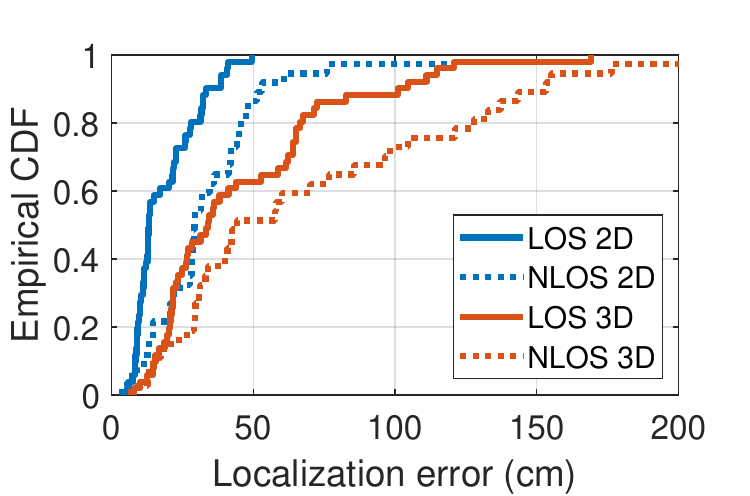}
    \includegraphics[width=1.6in]{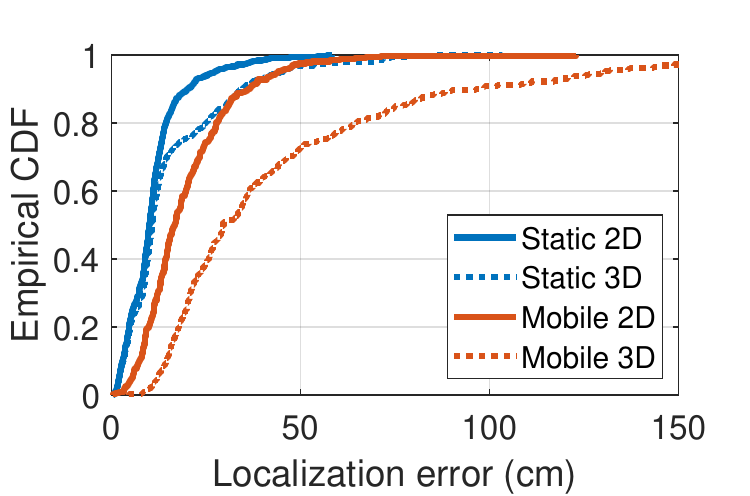}
    \caption{(a) LOS and NLOS localization errors. (b) Static vs Mobile nodes localization errors}
    \label{fig:los_vs_nlos}
\end{figure}

\textbf{LOS vs NLOS:} 
Figure \ref{fig:los_vs_nlos}(a) presents the CDF of the 2D and 3D localization errors in both line-of-sight and non-line-of-sight scenarios.
In the NLOS scenario, there is a slight increase in localization error compared to the LOS scenario. 
This increase is primarily attributed to the higher noise levels in the received signals at the nodes due to obstructed signal paths and reflections. 
Specifically, in the 2D scenario, the median localization error rises from $13cm$ in LOS to $28cm$ in NLOS.
Similarly, in the 3D scenario, the median localization error increases from $31cm$ in LOS to $39cm$ in NLOS.
\new

\textbf{Static vs Mobile nodes:} To assess the impact of node mobility on the performance of \name, we conduct experiments to compare static and mobile node scenarios.
Figure \ref{fig:los_vs_nlos}(b) presents the CDF of 2D and 3D localization errors in both static and mobile node configurations.
In the mobile scenario, the system experiences a slight degradation in performance due to varying environment and node vibrations. 
In the 2D scenario, the median localization error increases from a low of $6 cm$ in the static setup to $13 cm$ in the mobile scenario. Similarly, in the 3D scenario, the median localization error rises from $6.5 cm$ in the static scenario to $26 cm$ in the mobile scenario.

\subsection{Orientation Estimation accuracy}
\begin{figure}[htb]
    \includegraphics[width=1.6in]{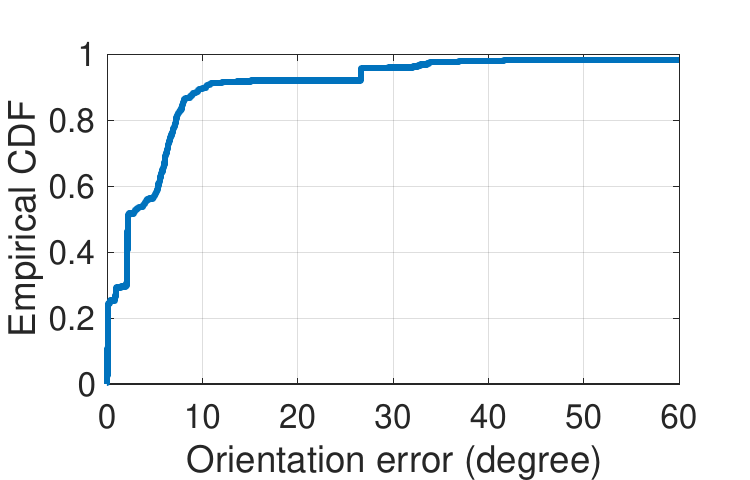}
    \includegraphics[width=1.6in]{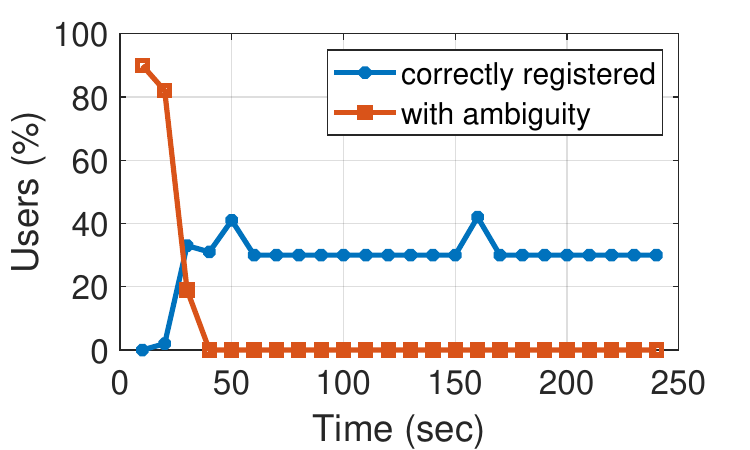}
    \caption{(a) CDF orientation errors. (b) User registration over time.}
    \label{fig:overall_orientation}
\end{figure}

\vspace{-0.07in}
Figure \ref{fig:overall_orientation}(a) presents the CDF of the error in orientation estimation results.
It is highly important for our system as these orientations play a critical role in transforming AoAs from the local frame of reference of the nodes to the global frame of reference, which is then used  localization of nodes within a 3D space.
Figure \ref{fig:overall_orientation}(a) shows that \name consistently estimate accurate node's orientation with a median error of 4 degrees which is competitive with an inertial sensor.

\subsection{Virtual Anchor Registration}
\vspace{-0.07in}
\name opportunistically integrate virtual anchors by computing the correlation of trajectories which facilitate the precise localization of nodes within a 3D space.
In Figure \ref{fig:User_registeration}, we demonstrate the selection of an optimal trajectory matching threshold, which maximizes user registration while minimizing registration time. 
Figure \ref{fig:overall_orientation}(b) depicts the dynamic process of user registration over time. 
It is evident that, within the first 40 seconds, \name successfully registers $40\%$ of the users, resolving any ambiguity in node locations. 



\subsection{Ranging and AoA Performance}
\vspace{-0.07in}
The localization errors inherently depend on the raging and AoA accuracies.
We report the ranging and AoA estimation performance of the UWB sensor used in our implementation.

While our system can be applied to any technology, here in the implementation we have used the UWB radio to get the distance and AoA error. So in this section we report the accuracy of the ranging and angle estimation accuracy.

\begin{figure}[tb]
    \centering
    \includegraphics[width=1.6in]{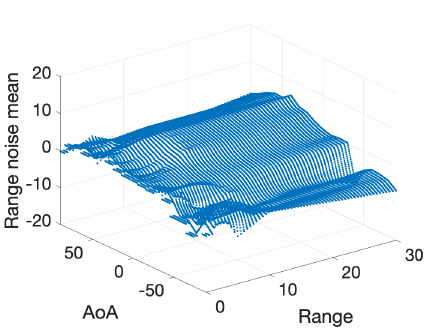}
    \includegraphics[width=1.6in]{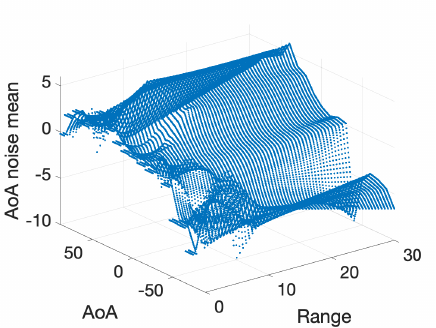}
    \caption{Ranging and AoA errors for varying distance and angles.}
    \label{fig:micro_ranging_aoa}
\end{figure}

Figure \ref{fig:micro_ranging_aoa} shows the ranging and AoA performance of our node for varying distances and incoming angles. We see that there are biases in certain angles as it is likely due to multipath. But still the error remain within $\pm 5^{\circ}$.

\section{Discussion}
\vspace{-0.1in}
\textbf{Fusing with Visual Inertial Odometry:} \name gives an option to opportunistically use VIO when available. While \name gives accurate 3D locations and orientations at slower intervals. The VIO can give a 30Hz odometry.
Each node gets a fine grained 3D tracking using on device VIO. We fuse it to get a more fine grained trajectory.
Combining UWB+VIO with Kalman or Bayesian Filters.
\new

\textbf{Cases when the subgraphs don't come in vicinity for a long time:} 
While this case is taken care of by our critical edges sampling method explained in Section 2.2. The subgraphs will accumulate drift errors if they rely on IMU for a long period of time. One solution is to trigger the bootstrap from start as the start and look for all possible edges to measure. In applications where the anchor infrastructure is available will instantly solve this problem since the subgraphs have a higher probability of having one of the static anchor in its network. Therefore pegging the relative subgraphs relative to each other.
\new

\textbf{No practical limitation of angle field-of-view:} New advancement in engineering these sensors can come up with better $360^\circ$ field-of-view sensors where someone could use larger array size or more antennas to cover the range. In such cases the Spanning tree algorithm could assume a binary case of either a range+angle edge or no edge. It could also account for different noise distributions for different incoming angle. This does not require any major modification and \name will handle it as well to give the optimal spanning tree.
\new

\textbf{Assumption of initial locations:} We assume that for any subsequent time iteration the system will have the previous time iteration locations and orientations of the nodes in network. While this assumption hold true most of the time, it may not be possible to create a complete graph in the initial t=0 instant or the system may not have enough time to create the first dense graph. This kind of scenario depends on the application where the system in deployed in and we leave this part for future work.
\new

\textbf{Other RF modalities:} Our works is basically a framework that can be applied to any modality and not just UWB and any RF, or acoustic signals are also possible since they have the capability to measure the pairwise ranges \cite{liu2020survey, wang2017research, liu2012push, peng2007beepbeep}.
\new
\section{Related Work}
\label{sec:related_work}
\vspace{-0.1in}
Over the past few decades, indoor positioning has attracted significant attention, and numerous studies have been conducted in this area. 
Existing work on indoor positioning can be broadly classified into two categories: infrastructure-based systems and infrastructure-free systems. 
In this section, we will provide an overview of related work in both categories.
\new

\textbf{Infrastructure-based systems:} 
Motion tracking cameras such as OptiTrack \cite{optitrack} and Vicon \cite{vicon} are used to estimate the locations of several users simultaneously.
These systems are often expensive and limited to the small pre-defined area of operation.
In addition to motion cameras, there are passive markers such as ARTags \cite{klopschitz2007automatic, mulloni2011handheld} and AprilTags \cite{wang2016apriltag, kallwies2020determining, zhao2020relative} which are commonly used in AR technology to precisely localize various users and objects in the environment. 
These markers can be localized precisely using only a camera and do not demand high computational power. 
However, these tags are only functional when the camera has a clear view of the tag, which means a large number of tags must be deployed in an environment to achieve broader coverage.
Apart from camera-based solution, beacon-based solutions utilize UWB \cite{nguyen2021range, olsson2014cooperative, shule2020uwb}, Bluetooth \cite{vcapkun2002gps, shao2018marble, lazik2015alps}, and ultrasound \cite{gomez2013indoor, lazik2015alps, li2018automatic} ranging for accurate localization.
These methods are often combined with odometry from either an IMU or VIO \cite{olsson2014cooperative, dhekne2019trackio, gentner2017simultaneous, liu2017cooperative, rajagopal2019improving, song2019uwb, wang2017ultra} to refine the location output and decrease the number of beacons necessary to cover the whole environment.
Although these solutions can localize multiple users simultaneously, infrastructure-based systems are fundamentally limited to the prepared environments where systems have already been installed.
\name adopts a strategy similar to these solutions however, instead of depending on stationary beacons in the environment, it leverages peer-to-peer UWB ranges and angles among multiple users which eliminates the need for any infrastructure.
\new

\textbf{Infrastructure-free systems:}
The idea of infrastructure-free localization is to determine the relative positions between users in a common coordinate system.
The absolute localization in a global coordinate system inherently requires some infrastructure/reference which is infeasible in many AR scenarios.
The idea of relative localization has been explored in sensor networks to localize static as well as moving nodes.
In sensor networks, relative localization has been studied to locate both stationary \cite{nagpal2003organizing, savvides2001dynamic} and mobile \cite{eren2004rigidity, moore2004robust, rad2011cooperative} nodes. 
Infrastructure-free localization has also been studied in robotics for the localization of drones or robots, using visual object detection \cite{nguyen2020vision, walter2019uvdar, ziegler2021distributed} or a combination of odometry and distance measurements \cite{bai2020high, cao2021relative, guo2019ultra, guo2017ultra, li2020relative, xu2020decentralized, xu2022omni}. 
These techniques primarily use windowed graph-based optimization or online filtering to merge sensor data. Recent research \cite{xu2020decentralized, xu2022omni} has employed visual-inertial features and UWB ranges to solve windowed optimization problems for localization. 
However, most of these methods have only been evaluated in small environments with limited trajectories, where devices are mostly in line-of-sight (LOS) which makes them impractical for tracking over large areas.
\new

The work that is most closely related is Cappella \cite{miller2022cappella}, which proposes an infrastructure-free positioning system for multi-user AR applications. 
Cappella uses motion estimates from VIO and range measurements from UWB between users to achieve precise localization in a relative coordinate system. 
In contrast, \name utilizes both range and angle information from UWB to achieve 6DOF location estimation. 
Unlike Cappella, \name does not heavily rely on VIO, which can fail in low lighting conditions. 
Additionally, \name achieves localization in a single shot and does not depend on past motion estimates, which helps to prevent error accumulation. 
To accomplish \name, a new optimization algorithm for localization has been developed that jointly optimizes both range and angle measurements from UWB.
\new

\section{Conclusion}
\vspace{-0.1in}
This paper presents \name, a state-of-the-art system designed for 3D localization and orientation estimation for large networks.
Our novel optimization algorithm integrates both range and angle-of-arrival measurements for enhanced network topology estimation, demanding fewer edge measurements. As a result, \name improves the latency by 75\% without sacrificing accuracy, outperforming conventional range-only solutions.


\bibliographystyle{plain}
\bibliography{references}
\balance

\end{document}